\documentclass[conference]{IEEEtran}
\usepackage{subfigure}
\usepackage{lineno,hyperref}
\usepackage[table,xcdraw]{xcolor}
\usepackage{amsmath}
\usepackage{cases}
\usepackage{graphicx}
\usepackage[margin=1in]{geometry}
\usepackage{color}
\usepackage{array}
\usepackage{multirow}
\newcolumntype{P}[1]{>{\centering\arraybackslash}p{#1}}
\usepackage{caption}
\begin{document}
\title{A note on power allocation for optimal capacity}
\author{Shravan Mohan\\ 17-004, Mantri Residency, Bannerghatta Main Road, Bangalore.  
}
\newgeometry{top=1in,bottom=0.75in,right=0.75in,left=0.75in}
\maketitle

\begin{abstract}
The problems of determining the optimal power allocation, within maximum power bounds, to (i) maximize the minimum Shannon capacity, and (ii) minimize the weighted latency are considered. In the first case, the global optima can be achieved in polynomial time by solving a sequence of linear programs (LP). In the second case, the original non-convex problem is replaced by a convex surrogate (a geometric program), using a functional approximation. Since the approximation error is relatively low, the optima of the surrogate is close to the global optimal point of the original problem. In either cases, there is no assumption on the SINR range. The use of LPs and geometric programming make the proposed algorithms numerically efficient. Computations are provided for corroboration. 
\end{abstract}
\section{Introduction}
Consider a system with $M$ links, which can interfere with each other. Suppose that the channel gain matrix is given by $G$, and that the maximum power that can be transmitted from the $i^{\rm th}$ link is given by $P_{i}$. Let $p_i$ be the power transmitted by the $i^{\rm th}$ transmitter and $\sigma^2_i$ be the noise power at the $i^{\rm th}$ receiver. In that case, the signal to interference and noise ratio at the receiver for the $i$ link is given by (see \cite{chiang2007power}):
\begin{equation}
    \begin{array}{l}
         \displaystyle \mbox{SINR}_i = \frac{G_{i,i} p_i}{\displaystyle\sum_{j\neq i} G_{j,i}p_j + \sigma^2_i}. 
    \end{array}
\end{equation}
The Shannon capacity of link $i$ is then given by (see \cite{cover1999elements}):
\begin{equation}
    C_i = \log\left( 1 + \mbox{SINR}_i \right).
\end{equation}
Typically, the objective in such problems is to optimize a metric such as: (i) maximizing the capacity of a particular link, (ii) minimizing the total power consumed, (iii) maximizing a weighted sum of link capacities, (iv) maximizing the minimum capacity across links, or (v) minimizing a weighted sum of latencies experienced in the links, among others (see \cite{chiang2007power}). It is well known that the problems (ii) and (iii) can be easily solved using convex methods, while (iii) poses a tough challenge \cite{chiang2007power}. In this paper, the problems of (iv) maximizing the minimum capacity across links, and (v) minimizing a weighted sum of latencies, are considered.  These shall be referred to as case (1) and case (2), respectively.

\begin{figure}[t]
    \centering{
    \includegraphics[width=3.5in]{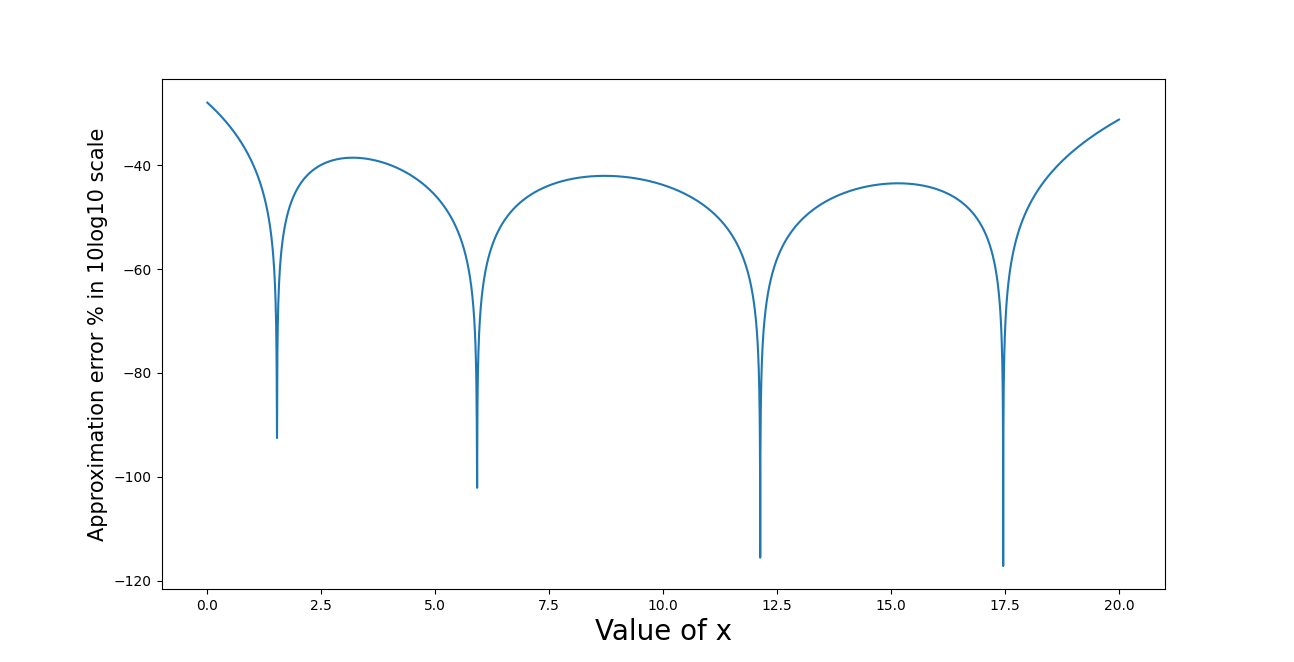}
    \caption{The approximation error percentage when $(e^x-1)^{\frac{1}{20}}$ is approximated as $\left(9.959\times 10^{-1}x^{\frac{1}{20}} + 2.859\times 10^{-2}x^{\frac{21}{20}} + 1.817\times 10^{-3}x^{\frac{41}{20}} \right.$ $ \left.+ 4.874\times 10^{-6}x^{\frac{61}{20}}\right)$.}
    \label{fig:approx_error}
    \vspace{-0.7cm}}
\end{figure}

The problem of optimal power allocation has received considerable attention from researchers. In \cite{chiang2007power}, the authors use geometric programming (see \cite{boyd2007tutorial}) to solve the whole suite of problems (by approximating $\log(1+\mbox{SINR})\approx \log(\mbox{SINR})$) under the assumption of high SINR. However, in the low/medium SINR range, the approximation fails and they resort a successive approximation method to attain a local optima. In \cite{qian2009mapel}, the authors propose a method based on monotonic optimization, where the constraint set is repeatedly approximated using normal sets. Since the cost functions are monotonic too, the maximum at each step is found out by evaluation at the boundary vertices. However, achieving global optima can be quite time consuming (see \cite{kha2011fast}). In \cite{kha2011fast}, the authors formulate the problems of weighted capacity maximization, and that of maximizing minimum of link capacities, as DC programming problems. These are solved using standard routines akin to SQP (see \cite{boyd2004convex}). Although their method is not guaranteed to lead to the global optima, it does quite well in practice with impressive convergence rates.

In contrast to the papers mentioned above, the method proposed in this paper (i) yields the global optima using linear programming for case (1), and (ii) in case (2), can get close to global optima of the original non-convex problem depending on the posynomial approximation error of $\left(e^x-1\right)^{\frac{1}{20}}$, over the range $[0,20]$. These aspects shall be discussed in detail.
\section{Algorithms}
The constraints in the two optimization problems considered in this paper would be (a) the maximum transmit power, and (b) the minimum desirable rate in the links. In case (1), only (a) shall be considered, while in case (2), both shall be considered. In addition to this, a constraint on outage probability can also be included \cite{chiang2007power}. However, this will not be used in this paper for ease of exposition. Note that in case (2), the feasibility of a problem with a given set of constraints can be as ascertained by solving a LP with those constraints and an arbitrary linear cost function \cite{chiang2007power}. 
\subsection{Maximizing minimum capacity across links}
\begin{figure*}[h!]
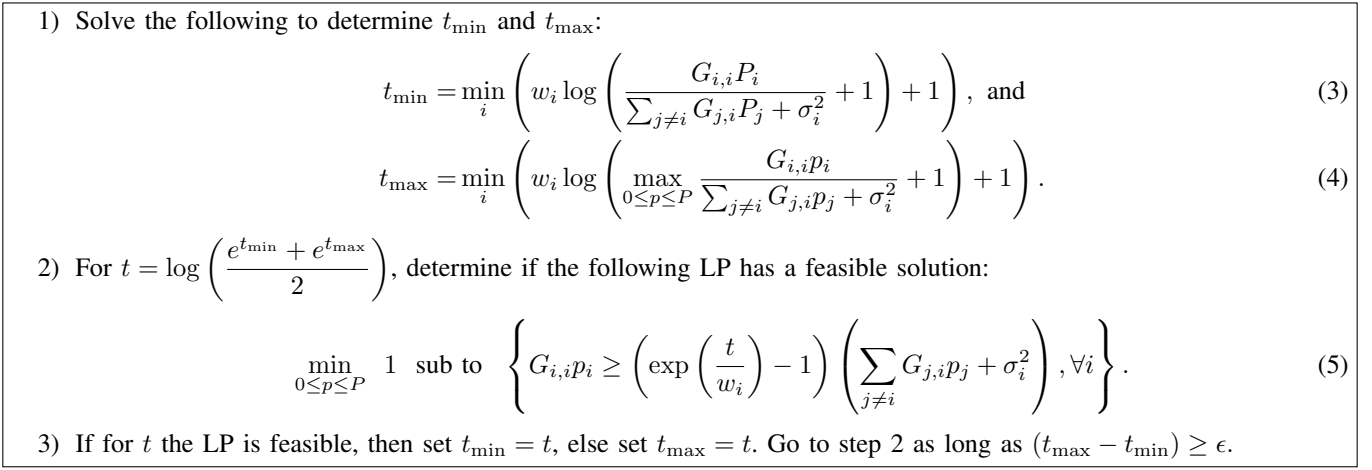

\centering
\fbox{\begin{minipage}{7in}
\begin{enumerate}
    \item {Solve the following to determine $t_{\min}$ and $t_{\max}$:
    \begin{align}
        t_{\min} =&\displaystyle \min_{i} \left(w_i\log\left(  \frac{G_{i,i}P_i}{\sum_{j\neq i}G_{j,i}P_j + \sigma^2_i} + 1\right) + 1\right), \mbox{~and}\\
        t_{\max} = &\displaystyle \min_{i} \left(w_i\log\left( \max_{0\leq p\leq P} \frac{G_{i,i}p_i}{\sum_{j\neq i}G_{j,i}p_j + \sigma^2_i}+1\right) + 1\right).
    \end{align}
    }
    \item { For $\displaystyle t = \log\left(\frac{e^{t_{\min}} + e^{t_{\max}}}{2}\right)$, determine if the following LP has a feasible solution:
    \begin{align}
        \min_{0\leq p\leq P} &~~1~~
        \mbox{sub to}~~
        \left\{G_{i,i}p_i \geq \left(\exp\left( \frac{t}{w_i}\right)-1 \right) \left(\sum_{j\neq i}G_{j,i}p_j + \sigma^2_i\right), \forall i\right\}.
    \end{align}
    }
    \item {If for $t$ the LP is feasible, then set $t_{\min} = t$, else set $t_{\max} = t$. Go to step 2 as long as $\left(t_{\max}-t_{\min}\right) \geq \epsilon.$
    }
\end{enumerate}
\end{minipage}}
\caption{The Max-Min Algorithm}
\label{fig:minmaxalgo}
\end{figure*}
\begin{table*}[h!]
\centering
\caption{Parameters and results pertaining to two examples for maximization of minimum capacity.}
\resizebox{6.9in}{!}{\begin{tabular}{|c|c|c|c|c|c|}
\hline
Gain Matrix & P (mW) & $\sigma^2 ~(\mu$W) & w & $p^*$ (mW) & Rates Achieved \\ \hline
G  = $\begin{bmatrix}
0.4310 & 0.0002 & 0.2605 & 0.0039 \\
0.0002 & 0.3018 & 0.0008 & 0.0054 \\
0.0129 & 0.0005 & 0.4266 & 0.1007 \\
0.0011 & 0.0031 & 0.0099 & 0.0634
    \end{bmatrix}$         & $[0.7, 0.8, 0.9, 1]$   & $[0.1, 0.1, 0.1, 0.1]$   & $[1/6, 1/6, 1/3, 1/3]$  &   $\color{blue}[0.1138, 0.1271, 0.2362, 0.9998]$     &   $\color{blue}[3.6425, 3.6425, 1.8212, 1.8212]$            \\ \hline
G  = $\begin{bmatrix}
0.4398, 0.0013, 0.0059, 0.0130, 0.0030, 0.0018, 0.0078, 0.0044, 0.0213, 0.0006\\
0.0161, 0.2252, 0.0108, 0.0020, 0.0030, 0.0088, 0.0043, 0.0457, 0.0093, 0.0100\\
0.0043, 0.0065, 0.1470, 0.7860, 0.0600, 0.0197, 0.0100, 0.0240, 0.0022, 0.0200\\
0.0063, 0.0516, 0.0074, 0.3284, 0.1630, 0.0276, 0.0182, 0.0365, 0.0014, 0.0294\\
0.0225, 0.0297, 0.0016, 0.0033, 0.4220, 0.0352, 0.0169, 0.0148, 0.0031, 0.0163\\
0.0599, 0.0683, 0.0112, 0.0062, 0.0470, 0.1146, 0.0310, 0.0169, 0.0045, 0.0178\\
0.0149, 0.0580, 0.0115, 0.0667, 0.0730, 0.0017, 0.3600, 0.0009, 0.0007, 0.0036\\
0.0208, 0.0065, 0.0033, 0.0017, 0.0460, 0.0061, 0.0233, 0.5481, 0.0198, 0.0775\\
0.0025, 0.0038, 0.0397, 0.0010, 0.0330, 0.0027, 0.0289, 0.0955, 0.6540, 0.0527\\
0.0004, 0.0099, 0.0378, 0.0153, 0.0634, 0.0047, 0.0213, 0.0006, 0.0086, 0.1695
    \end{bmatrix}$           &  \begin{tabular}[c]{@{}c@{}}[0.5, 0.55, 0.6, 0.65, 0.7, \\ 0.75, 0.8, 0.85, 0.9, 0.95]\end{tabular} &   \begin{tabular}[c]{@{}c@{}}[0.1, 0.1, 0.1, 0.1, 0.1,\\ 0.1, 0.1, 0.1, 0.1, 0.1]\end{tabular}  & \begin{tabular}[c]{@{}c@{}}[2/15, 2/15, 2/15, 2/15, 2/15,\\ 2/15, 1/15, 1/15, 1/15, 1/15]\end{tabular} &  \color{blue}   \begin{tabular}[c]{@{}c@{}}[0.0935, 0.3589, 0.2907, 0.6479, 0.4387,\\ 0.3338, 0.3722  , 0.2682, 0.0759, 0.8682]\end{tabular}  &    \color{blue}     \begin{tabular}[c]{@{}c@{}}[0.8321, 0.8321, 0.8321, 0.8321, 0.8321,\\ 0.8321, 1.6642, 1.6642, 1.6642, 1.6642]\end{tabular}       \\ \hline
\end{tabular}}
\label{tab:min_max}
\end{table*}
Consider the following optimization problem
\begin{equation}
    \begin{array}{l}
         \max ~\min ~w_i C_i\\
         \mbox{sub to}\\
         ~~~~~~~0\leq p_i\leq P_i, \forall i.
    \end{array}
\end{equation}
The optimization problem as such is non-convex and hence hard to solve. However, a transformation of this problem leads to a procedure to obtain the globally optimal solution. So, for a given $t$, consider the following feasibility problem
\begin{equation}
    \begin{array}{l}
        \displaystyle ~~~~~~~~~~~\mbox{Find~~} p \mbox{~~such that}\\
         \displaystyle G_{i,i}p_i \geq \left(e^{ \frac{t}{w_i}} -1\right)\left(\sum_{j\neq i}G_{j,i}p_j + \sigma^2_i\right), \forall i \\
         ~~~~~~~~~~~~0\leq p_i\leq P_i, \forall i.
    \end{array}
\end{equation}
Naturally, the optimal solution to the original problem is the largest $t$ for which the above feasibility problem has a solution. Note that the existence of such a feasible point, for any given $t$, can be ascertained using the following  LP:
\begin{equation}
    \begin{array}{l}
        \displaystyle ~~~~~~~~~~~~\min ~~\sum_i p_i\\
         \displaystyle G_{i,i}p_i \geq \left(e^{ \frac{t}{w_i}} -1\right)\left(\sum_{j\neq i}G_{j,i}p_j + \sigma^2_i\right), \forall i \\
         ~~~~~~~~~~~~0\leq p_i\leq P_i, \forall i.
    \end{array}
\end{equation}
\begin{figure*}[h!]
\centering
\fbox{\begin{minipage}{7in}
\begin{enumerate}
    \item {Set $t_{\min}=0$. Solve the following to determine $t_{\max}$:
    \begin{align}
        t_{i, \max} = &\displaystyle  \max_{0\leq p\leq P} \frac{G_{i,i}p_i}{\sum_{j\neq i}G_{j,i}p_j + \sigma^2_i} \mbox{~s.t.} \left\{\frac{G_{i,i}p_i}{\sum_{j\neq i}G_{j,i}p_j + \sigma^2_i} + 1 \geq e^{r_i}, \forall i\right\} 
        \mbox{~\&~} t_{\max} = \displaystyle \max_{i} \lceil w_i\log\left(t_{i, \max}+1\right)\rceil
    \end{align}
    }
    \item {Set $\displaystyle d = \left[\frac{1}{t_{\max}}, \frac{2}{t_{\max}}, \cdots, \frac{t_{\max}-1}{t_{\max}}, 1, 2\right]$}
    \item { Solve the following convex optimization problem to determine coefficients:
    \begin{align}
        \min_{c\geq 0}~~ c^\top A c + b^\top c,
    \end{align}
    where $\displaystyle A_{i,j} = \left(1 + I(i\neq j)\right)\int^{t_{\max}}_{t_{\min}}x^{d_i+d_j}dx$,  and $\displaystyle b_i = -2\int^{t_{\max}}_{t_{\min}}\left(e^{x}-1\right)^{\frac{1}{t_{\max}}}x^{d_i}dx$.
    }
    \item {Solve the following Geometric Program:
    \begin{align}
        &\min_{p, t, z} ~~~~~~\sum_i ~~\frac{w_i}{t_i}\\
        &\mbox{sub to}\nonumber \\
        &~~~~~~G_{i,i}p_i \geq \left(e^{r_i} -1\right)\left(\sum_{j\neq i}G_{j,i}p_j + \sigma^2_i\right), \forall i,\\
        &~~~~~~\left(G_{i,i}p_i\right)^{\frac{1}{t_{\max}}} \geq \left(\sum^{t_{\max} + 2}_{k=1}c_kt_i^{d_k}\right)z_i^{\frac{1}{t_{\max}}}, \forall i,\\
        &~~~~~~z_i\geq \left(\sum_{j\neq i}G_{j,i}p_j + \sigma^2_i\right), \forall i,\\
        &~~~~~~ 0\leq p\leq P
    \end{align}
    }
    \item {If the above geometric program yields a feasible solution, return it.}
\end{enumerate}
\end{minipage}}
\caption{The Latency Minimization Algorithm}
\vspace*{-0.4cm}
\label{fig:latencyalgo}
\end{figure*}

\begin{table*}[h!]
\centering
\caption{Parameters and results pertaining to two examples for minimization of latencies.}
\resizebox{6.9in}{!}{\begin{tabular}{|c|c|c|c|c|c|c|}
\hline
Gain Matrix & P (mW) & $\sigma^2 ~(\mu$W) & w & r & $p^*$ (mW) & Rates Achieved \\ \hline
G  = $\begin{bmatrix}
0.4310 & 0.0002 & 0.2605 & 0.0039 \\
0.0002 & 0.3018 & 0.0008 & 0.0054 \\
0.0129 & 0.0005 & 0.4266 & 0.1007 \\
0.0011 & 0.0031 & 0.0099 & 0.0634
    \end{bmatrix}$         & $[0.7, 0.8, 0.9, 1]$  & $[0.1, 0.1, 0.1, 0.1]$ &  $[1/6, 1/6, 1/3, 1/3]$     &   $[1, 1, 1, 1]$ & $\color{blue}[0.0285,   0.3240,   0.1367,   1.0000]$   & $\color{blue}[ 1.6203,   3.4298,   1.4585,   1.6154]$               \\ \hline
G  = $\begin{bmatrix}
0.4398, 0.0013, 0.0059, 0.0130, 0.0030, 0.0018, 0.0078, 0.0044, 0.0213, 0.0006\\
0.0161, 0.2252, 0.0108, 0.0020, 0.0030, 0.0088, 0.0043, 0.0457, 0.0093, 0.0100\\
0.0043, 0.0065, 0.1470, 0.7860, 0.0600, 0.0197, 0.0100, 0.0240, 0.0022, 0.0200\\
0.0063, 0.0516, 0.0074, 0.3284, 0.1630, 0.0276, 0.0182, 0.0365, 0.0014, 0.0294\\
0.0225, 0.0297, 0.0016, 0.0033, 0.4220, 0.0352, 0.0169, 0.0148, 0.0031, 0.0163\\
0.0599, 0.0683, 0.0112, 0.0062, 0.0470, 0.1146, 0.0310, 0.0169, 0.0045, 0.0178\\
0.0149, 0.0580, 0.0115, 0.0667, 0.0730, 0.0017, 0.3600, 0.0009, 0.0007, 0.0036\\
0.0208, 0.0065, 0.0033, 0.0017, 0.0460, 0.0061, 0.0233, 0.5481, 0.0198, 0.0775\\
0.0025, 0.0038, 0.0397, 0.0010, 0.0330, 0.0027, 0.0289, 0.0955, 0.6540, 0.0527\\
0.0004, 0.0099, 0.0378, 0.0153, 0.0634, 0.0047, 0.0213, 0.0006, 0.0086, 0.1695
    \end{bmatrix}$           & \begin{tabular}[c]{@{}c@{}}[1, 1, 1, 1, 1,\\ 1, 1, 1, 1, 1]\end{tabular}   &  \begin{tabular}[c]{@{}c@{}}[1, 1, 1, 1, 1,\\ 1, 1, 1, 1, 1]\end{tabular}     &  \begin{tabular}[c]{@{}c@{}}[2/15, 2/15, 2/15, 2/15, 2/15,\\ 2/15, 1/15, 1/15, 1/15, 1/15]\end{tabular}  & \begin{tabular}[c]{@{}c@{}}[0.5, 0.5, 0.5, 0.5, 0.5,\\ 0.5, 0.5, 0.5, 0.5, 0.5]\end{tabular}  &  \color{blue} \begin{tabular}[c]{@{}c@{}}[0.8829,   1.0000,   0.3399,   0.7913,   0.8628,\\  0.7530,   0.3534,   0.3466,   0.1580,   0.5020]\end{tabular}    &   \color{blue}  \begin{tabular}[c]{@{}c@{}}[1.5767,   0.9145,   0.5830,   0.5928,   0.8533,\\ 0.7601,   0.8730,   0.9060,   1.1420,   0.5912]\end{tabular}           \\ \hline
\end{tabular}}
\vspace*{-0.5cm}
\label{tab:latency}
\end{table*}
Now, if there is no feasible point satisfying the above linear inequalities for a given $t$, then for any $\delta>0$, there is no feasible point for $t+\delta$. This can be proved by contradiction. Suppose there was a feasible point for $t+\delta$, then the same would be a feasible point for $t$ too, a contradiction to the earlier assumption. This implies that the largest $t$ where a feasible solution exists can be determined using bi-section search, between 0 and $t_{\max}$, at which there is no feasible point. The determination of the quantity $t_{\max}$, in turn, requires the maximization of each of the SINRs. For maximizing any one of the SINRs, one can use Dinkelbach's algorithm \cite{boyd2004convex}. That is, for the link $i$, start with $\lambda_i = 1$, and solve the following convex optimization problem:
\begin{align}
    \max ~~G_{i,i}p_i - \lambda_i \sum_{j\neq i}G_{j,i}p_j~~
    \mbox{sub to}~~ 0\leq p \leq P.
\end{align}
Set $\displaystyle \lambda_i = \frac{G_{i,i}p^*_i}{\sum_{j\neq i}G_{j,i}p^*_j}$, where $p^*$ is the optimal solution to the above problem. Solve the problem with this new $\lambda_i$ and repeat till convergence. Applying Dinkelbach's algorithm to all the links, one can choose $t_{\max} = \min_i \{\lambda_i + 1\}$. It is clear that there is no feasible point at $t_{\max}$.  With $t_{\max}$, the bi-section search can be carried out as usual (see \cite{boyd2004convex}); the complete algorithm is presented in \ref{fig:minmaxalgo}. Thus, the optimal power allocation that leads to the max-min data rate can be determined in polynomial time (polynomial in the number of links and $\log(\epsilon)$, where $\epsilon$ is the numerical tolerance; see \cite{boyd2004convex}).

\subsection{Minimizing weighted sum of latencies}
The optimal power allocation for weighted data rate maximization, which is stated below, is more challenging a problem than the max-min data rate problem:
\begin{equation}
    \begin{array}{l}
         \min ~~~~~\displaystyle \sum_{k=1}^M w_i/R_i\\
         \mbox{sub to}\\
         ~~~~~~~\displaystyle R_i = \log \left(1+SINR_i\right) \geq r_i, \forall i\\
         ~~~~~~~\displaystyle 0\leq p_i\leq P_i, \forall i.
    \end{array}
\end{equation}
The solution approach proposed here uses geometric programming, by replacing some terms with appropriate posynomials. It will be assumed that the maximum achievable SINR is $2^{20}$ (or equivalently $\tilde 60$dB), which is excellent signal quality in practice. Now, consider the constraint
\begin{align}
    G_{i,i}p_i \geq \left(e^{t_i} -1\right)\left(\sum_{j\neq i}G_{j,i}p_j + \sigma^2_i\right)
\end{align}
for some $i$. A naive way to make this constraint amenable to geometric programming is to approximate $e^{t_i} -1$ as $\sum^{K}_{k=1}\frac{t_i^k}{k!}$, for some large $K$. However, if $t_i$ were in the range $[0,20]$ (as per the earlier assumption), one would need $K=20$ for a reasonably good approximation. However, the coefficient $\frac{1}{20!}$ becomes too small for numerical purposes. A possible way to get around this is to change it to:
\begin{align}  
\left(G_{i,i}p_i\right)^{\frac{1}{20}} \geq \left(e^{t_i} -1\right)^{\frac{1}{20}}\left(\sum_{j\neq i}G_{j,i}p_j + \sigma^2_i\right)^{\frac{1}{20}}.
\end{align}
If $\left(e^{t_i} -1\right)^{\frac{1}{20}}$ can be approximated well as $\sum^{K}_{k=1} c_k t^{d_k}$, for an appropriate choice rational exponents and positive coefficients, then the constraint becomes amenable to geometric programming. Intuitively, note that around $x=0$, the function $\left(e^{t_i} -1\right)^{\frac{1}{20}}$ behaves like $x^{\frac{1}{10}}$. And around $x=20$, the function behaves like $e^x$ behaves around $x=1$. So, one can expect a reasonably good approximation with lower order terms. Now, the Newton-Puiseux (see \cite{willis2003newton}) series guarantees the existence of a globally convergent fractional power series representation of the function in consideration. This series only includes terms with powers of the form $k+1/20, k\in Z$ and incidentally, the coefficient of the first four terms are positive. This already yields a good approximation over the desired range. However, one can compute a better approximation by determining the set of non-negative coefficients that minimize the total squared approximation error for the given terms in the posynomial. That is, by solving the following convex problem (for a particular choice of K, typically 4):
\begin{equation}
    \begin{array}{l}
         \displaystyle ~~~~~~~~\min_{c\geq 0}~~ c^\top A c + b^\top c, \mbox{~~where}\\
        \displaystyle A_{i,j} = \left(1 + I(i\neq j)\right)\int^{t_{\max}}_{t_{\min}}x^{d_i+d_j}dx, \mbox{~~and}\\
         \displaystyle ~~~~~~~b_i = -2\int^{t_{\max}}_{t_{\min}}\left(e^{x}-1\right)^{\frac{1}{t_{\max}}}x^{d_i}dx.
    \end{array}
\end{equation}
The approximation error percentage is shown (in log scale) in \ref{fig:approx_error}, which is certainly within acceptable range. With that, the optimization problem can be reformulated as the following geometric program:
\begin{equation}
    \begin{array}{l}
        \displaystyle \min_{p, t, z} ~~~~~~\sum_i ~~\frac{w_i}{t_i}\\
        \displaystyle \mbox{sub to} 
        \displaystyle ~~~~~~G_{i,i}p_i \geq \left(2^{r_i} -1\right)\left(\sum_{j\neq i}G_{j,i}p_j + \sigma^2_i\right), \forall i,\\
        \displaystyle ~~~~~~\left(G_{i,i}p_i\right)^{\frac{1}{20}} \geq \left(\sum^{K}_{k=1}c_kt_i^{d_k}\right)z_i^{\frac{1}{20}}, \forall i,\\
        \displaystyle ~~~~~~z_i\geq \left(\sum_{j\neq i}G_{j,i}p_j + \sigma^2_i\right), \forall i,
        \displaystyle ~~~ 0\leq p\leq P.
    \end{array}
\end{equation}
Therefore, the global optima of the above optimization problem can be found in polynomial time (polynomial in the number of links and $\log(\epsilon)$, where $\epsilon$ is the numerical tolerance) \cite{boyd2004convex}.

Note that at optimality, the set of constraints (13) and (14) involving the variables $z$ will be tight. That can be proved by contradiction. Firstly, since the variables $t$ appear in the denominator, for any $z^*$ at optima, each element of $t^*$ should be the maximum possible. Secondly, note that the set of constraints (13) and (14) are not mixed in $t$ or $z$. Thus, the set of constraints (13) must be tight at optimality. Now, if (14) were not all tight, then a reduction in one of the variables in $z$, say $z_i$, will lead to an increase in $t_i$, thereby leading to a smaller cost. A contradiction to the assumption of optimality. Thus, the set of constraints (14) must also be tight at 
optimality. The tightness of these constraints at optimality establishes the equivalence between the original non-convex optimization problem and the geometric program. Finally, since $\displaystyle (e^x-1)^{\frac{1}{20}}$ is approximated well by a posynomial, the global optima of the geometric program will be close to the global optima of the original non-convex problem. The complete algorithm is presented in \ref{fig:latencyalgo}. The algorithm will also detect infeasibility, if such a case occurs. 
\section{Computations}
Two examples are used from \cite{qian2009mapel}, and the parameters of the system and the outputs (from CVXPY \cite{diamond2016cvxpy} in case (1) and CVX using Octave \cite{boyd2004convex} in case (2)) are shown in \ref{tab:min_max} and \ref{tab:latency}. The parameters and results presented in \ref{tab:min_max} are related to maximizing the minimum weighted capacity, while those presented in \ref{tab:latency} relate to minimizing a weighted summation of latencies. As expected, in case (2), all constraints involving the auxiliary variables $z$ (that is, (13) and (14)) are tight at optimality for case (2). The obtained solution from the algorithm can be used as a starting point for a gradient based algorithm to further improve the cost in the original non-convex problem. However, it is has been observed that this polishing step yields only marginal improvements.
\section{Conclusions}
This paper presented a convex optimization approach to solving the power allocation problem for achieving optimal capacity. Two cases were considered, i.e., (i) maximize the minimum capacity among the links, and (ii) minimize the weighted sum of latencies across links. The global optima for the first problem was obtained in polynomial by solving a sequence of linear programs. In the second case, the original non-convex optimization problem was replaced by another convex (a geometric program), by using a functional approximation. Since the functional approximation error is relatively low, the optima of the surrogate would also be very close to the optima of the original problem. In both the cases, the method does not make any assumptions on the SINR range. Computations were provided for corroboration.   
\vspace*{-0.5cm}
\bibliographystyle{IEEEtran}
\bibliography{biblio}

\begin{thebibliography}{1}
\providecommand{\url}[1]{#1}
\csname url@samestyle\endcsname
\providecommand{\newblock}{\relax}
\providecommand{\bibinfo}[2]{#2}
\providecommand{\BIBentrySTDinterwordspacing}{\spaceskip=0pt\relax}
\providecommand{\BIBentryALTinterwordstretchfactor}{4}
\providecommand{\BIBentryALTinterwordspacing}{\spaceskip=\fontdimen2\font plus
\BIBentryALTinterwordstretchfactor\fontdimen3\font minus
  \fontdimen4\font\relax}
\providecommand{\BIBforeignlanguage}[2]{{%
\expandafter\ifx\csname l@#1\endcsname\relax
\typeout{** WARNING: IEEEtran.bst: No hyphenation pattern has been}%
\typeout{** loaded for the language `#1'. Using the pattern for}%
\typeout{** the default language instead.}%
\else
\language=\csname l@#1\endcsname
\fi
#2}}
\providecommand{\BIBdecl}{\relax}
\BIBdecl

\bibitem{chiang2007power}
M.~Chiang, C.~W. Tan, D.~P. Palomar, D.~O'neill, and D.~Julian, ``Power control
  by geometric programming,'' \emph{IEEE transactions on wireless
  communications}, vol.~6, no.~7, pp. 2640--2651, 2007.

\bibitem{cover1999elements}
T.~M. Cover, \emph{Elements of information theory}.\hskip 1em plus 0.5em minus
  0.4em\relax John Wiley \& Sons, 1999.

\bibitem{boyd2007tutorial}
S.~Boyd, S.-J. Kim, L.~Vandenberghe, and A.~Hassibi, ``A tutorial on geometric
  programming,'' \emph{Optimization and engineering}, vol.~8, no.~1, pp.
  67--127, 2007.

\bibitem{qian2009mapel}
L.~P. Qian, Y.~J. Zhang, and J.~Huang, ``Mapel: Achieving global optimality for
  a non-convex wireless power control problem,'' \emph{IEEE Transactions on
  Wireless Communications}, vol.~8, no.~3, pp. 1553--1563, 2009.

\bibitem{kha2011fast}
H.~H. Kha, H.~D. Tuan, and H.~H. Nguyen, ``Fast global optimal power allocation
  in wireless networks by local dc programming,'' \emph{IEEE Transactions on
  Wireless Communications}, vol.~11, no.~2, pp. 510--515, 2011.

\bibitem{boyd2004convex}
S.~Boyd and L.~Vandenberghe, \emph{Convex optimization}.\hskip 1em plus 0.5em
  minus 0.4em\relax Cambridge university press, 2004.

\bibitem{willis2003newton}
N.~J. Willis, ``Newton-puiseux algorithm,'' Ph.D. dissertation, Texas Tech
  University, 2003.

\bibitem{diamond2016cvxpy}
S.~Diamond and S.~Boyd, ``Cvxpy: A python-embedded modeling language for convex
  optimization,'' \emph{The Journal of Machine Learning Research}, vol.~17,
  no.~1, pp. 2909--2913, 2016.

\end{thebibliography}
\end{document}